# The location of the Fisher zeros and estimates of $y_T = 1/\nu$ are found for the Baxter-Wu model.

James L. Monroe*

Dept. of Physics, Pennsylvania State University, Beaver Campus

100 University Dr., Monaca PA 15061-2799

It is shown that the location of the Fisher zeros of the Baxter-Wu model, for two series of finite sized clusters, with spherical boundary conditions, is extremely simple. They lie on the unit circle in the complex $\mathrm{Sinh}[2\beta J_3]$ plane. This is the same location as the Fisher zeros of the Ising model with nearest neighbor interactions, $J_2$, on the square lattice have, with Brascamp-Kunz boundary conditions. The Baxter-Wu model is an Ising model with three site interactions, $J_3$, on the triangle lattice. From the leading Fisher zeros, using finite size scaling, accurate estimates of the critical exponent $1/\nu$ are obtained. Furthermore, using the imaginary parts of the leading zeros versus the real part of the leading zeros leads to different results similar the results of Janke and Kenna for the nearest neighbor, Ising model on the square lattice and extending this behavior to a multi-site interaction system.

## I. Introduction

C. N. Yang and T. D. Lee in the first of two seminal papers [1] established the importance of the zeros of the partition function in statistical mechanics, especially in regards to the occurrence of phase transitions. In the second of their papers [2] regarding the zeros of the partition function they examined the location of zeros in the complex $\exp[-2\beta h]$ plane, where h is the magnetic field and $\beta = 1/kT$, where k is the Boltzmann constant and T the temperature. They proved that for ferromagnetic pair interactions amongst a set of Ising spins, i.e., spins only taking on the two values ±1, that in the complex $\exp[-2\beta h]$ plane, the location of the zeros simply lie on the unit circle. Since then, zeros lying in some plane related to the magnetic field are generally known as Lee-Yang zeros.

Not long afterward Yang and Lee's introduction of the importance of the zeros of the partition function Michael Fisher [3] began the study of the zeros of the partition function involving some complex plane related to the temperature, T. One such plane is the complex $\exp[-2\beta J_2]$ plane, where $J_2$ is the interaction strength between a pair of Ising spins. Appropriately these zeros have become known as Fisher zeros.

*email address: jim5@psu.edu

While the location of the Lee-Yang zeros for the ferromagnetic pair interactions systems involving Ising spins is incredibly simple, this is the exception rather than the rule. This is especially true of Fisher zeros. In their review article, Bena, Droz, and Lipowski [4] give three reasons for the distribution of the Fisher zeros being more complicated than that of the Lee-Yang zeros. They then write "These elements explain the scarcity of exact analytical results on the Fisher zeros (as compared to the Yang-Lee zeros) …."

In the following I find for certain boundary conditions the location of the Fisher zeros for some finite sized clusters of the Baxter-Wu model and their location is both simple and similar to that of the nearest-neighbor interaction (hereafter n.n.), Ising model, on the square lattice. I also present results based on the finite size scaling (hereafter FSS) behavior of these zeros both getting accurate estimates for the correlation length critical exponent ν and finding that the behavior of the scaling of the imaginary and real parts of the complex zeros differ. This later characteristic is similar to what was recently found by Janke and Kenna [5] for the n.n., square lattice, Ising model with Brascamp-Kunz boundary conditions.

The Baxter-Wu model along with notation is presented in the following section. Then there are two sections containing results concerning the Fisher zeros of the model. The first of these two sections contain results regarding the location of the zeros for various finite sized systems and the second are results regarding critical and shift exponents found using the leading Fisher zeros and FSS. The final section contains conclusions along with some questions raised by these results.

**II. Baxter-Wu model and notation.**

The Baxter-Wu model is an Ising model on the triangle lattice with Hamiltonian

$$\mathcal{H}(\sigma) = -J_3 \sum_\Delta \sigma_i \, \sigma_j \sigma_k - h \sum_i \sigma_i \qquad (1)$$

where $J_3$ is the interactions strength of the three-site interactions involving Ising spins $\sigma_i$, $\sigma_j$, and $\sigma_k$, and $h$ is the magnetic field. The first sum is over all elementary triangles forming the lattice and the second is over all sites of the lattice. Here only the case of h = 0 will be considered. Results concerning the case with h ≠ 0 will be presented elsewhere. With this restriction, h = 0, this lattice spin system is one of a rather small number of lattice spin systems which have been solved exactly. In particular Baxter and Wu [6,7,8] showed the critical temperature to be exactly the same as the most famous of the exactly solved lattice spin systems, that of the n.n., square lattice, Ising system solved by Onsager [9]. The critical temperature for both models is 2/Log[1+√2]. Besides the critical temperature, the value of several critical exponents were found. Unlike the critical temperature, the critical exponents have values different from those found by Onsager for the n.n., ferromagnetic, Ising model on the square lattice. This is to be expected as the symmetries of both models differ, in particular the Baxter-Wu model lacks the up-down spin reversal symmetry of the pair interaction Ising models and is in the same universality class as the 4-state Potts model.

Knowing exact values for the critical temperature and critical exponents this model has occasionally been used as a test of various approximation methods, e.g., series expansion analysis [10, 11], real-space renormalization group [12], Monte Carlo renormalization group [13], and Wang-Landau sampling methods [14]. However, to the author's knowledge, there have been no studies of the zeros of the partition function and their implications for this model. For a summary of lattice spin systems for which the Fisher zeros have been studied see section 12 of [4].

For the Baxter-Wu model I calculate the exact partition function for a number of finite site clusters as is often done in trying to determine properties of the Fisher zeros. This started back when the Fisher zeros of the square lattice Ising model were first looked at [15, 16], and continues today [17, 18]. The partition function is

$$Z(\beta) = \sum_{\{\sigma\}} e^{-\beta \mathcal{H}(\sigma)} = \sum_{n=-N_b}^{N_b} \Omega(n) b^n \qquad (2)$$

where the first sum is over all configurations, denoted by {σ}. The second sum indicates the partition function written as a generalized polynomial in *b* where I have $b = \exp[\beta J_3]$. The partition function can be written in terms of $\exp[\pm\beta J_3]$, $\exp[\pm 2\beta J_3]$, or $\exp[\pm 4\beta J_3]$, the latter only in the case of periodic boundary conditions, as well as other quantities, some of which will be presented later. Obviously which variable one chooses to express the partition function clearly impacts the location of the zeros. Fisher's [3] original results used $\exp[-2\beta J_2]$ and it is in this complex plane that he showed the zeros lie on two intersecting circles in the thermodynamic limit.

### III. Location of the Fisher zeros for the Baxter-Wu model.

Perhaps the most obvious clusters of sites to use in investigating the zeros of the Baxter-Wu model are clusters of a square shape where each elementary square is divided by a single diagonal into two triangles creating a triangular lattice. I will denote the size of such clusters as *LxL* with *L* being the number of sites along any edge of the original square. This shape is generally studied using free, cylindrical, or toroidal boundary conditions. With any type of periodicity, it is critical to realize that the Baxter-Wu Hamiltonian creates a situation where there are three sublattices of importance and because of this only *LxL* clusters where *L* is divisible by 3 should be examined.

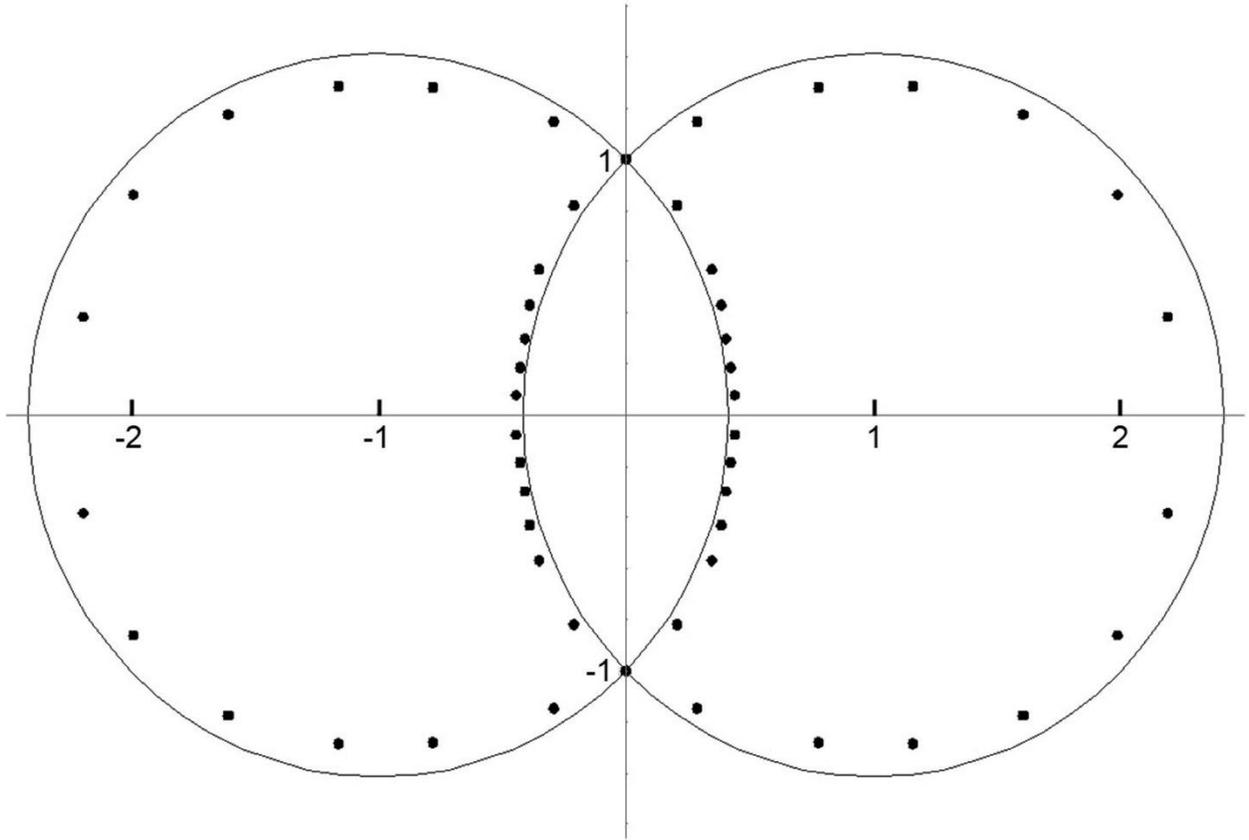

Figure 1. The location of the Fisher zeros for the 6x6 cluster with toroidal boundary conditions in the complex e$^{-2\beta J}$ plane with the two intersecting circles shown given by Eq. (3).

The location of the Fisher zeros for the 6x6 cluster with toroidal boundary conditions is shown in Fig. 1. The zeros lie close to, but not on, two intersecting circles similar to the case of the Fisher zeros of the n.n., square lattice, Ising model when finite clusters with toroidal boundary conditions are examined. In the thermodynamic limit Fisher [3] conjectured that the zeros lie on two interacting circles given by

$$e^{-2\beta J_2} = \pm 1 + \sqrt{2} e^{i\varphi} \tag{3}$$

where $J_2$ is the n.n. interaction strength. With the Baxter-Wu model I conjecture that with larger and larger square clusters and with any of the three commonly used boundary conditions the zeros move toward the two circles, and reach the two intersecting circles in the thermodynamic limit similar to what happens with similar clusters of sites for the n.n., Ising model on the square lattice.

Now with the n.n., square lattice, Ising model Brascamp and Kunz [19] established the fact that with two special boundary conditions of theirs, the zeros lie directly on the intersecting circles for all finite, as well as the infinite, systems. I do something similar for the case of the Baxter-Wu model by presenting two series of clusters and a special boundary condition where the zeros for all finite systems investigated lie exactly on the circles given in (3). Unlike the case considered by Brascamp and Kunz [19] there is no expression for the partition function that I have found where it can be shown that for any finite system with these boundary conditions and hence in the thermodynamic limit the zeros lie exactly on the circles. Rather, I present results for small finite systems where the partition function and hence the Fisher zeros can be directly calculated.

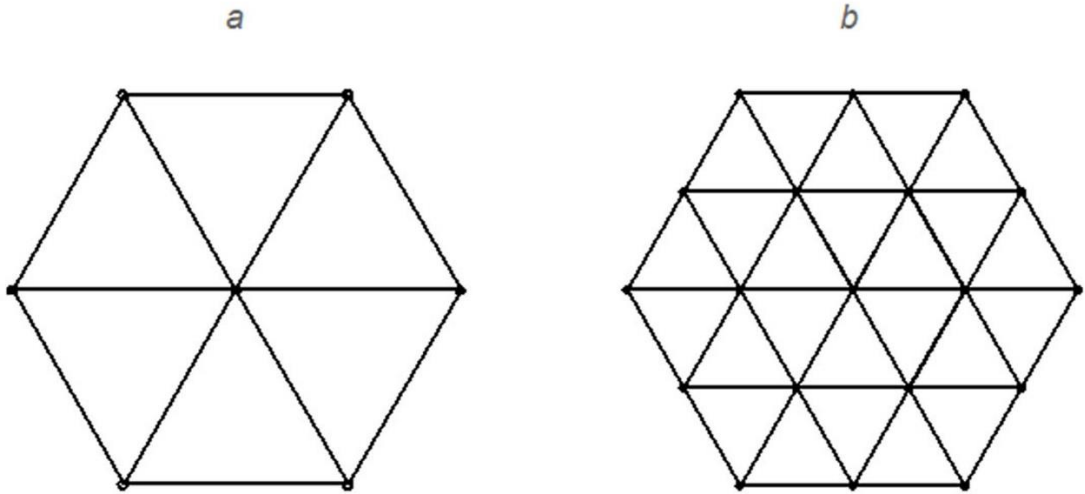

Fig. 2. The first and second clusters in the Series A.

The boundary conditions I employ with the Baxter-Wu model might be called "spherical" boundary conditions. They are easiest viewed in the following manner. One considers a hexagonal cluster consisting of a single central site and the six surrounding sites, see Fig. 2a. One has six three-site interactions with the sites around the corners of each elementary triangle. Now one should imagine this hexagonal cluster draped over the top half of a sphere thereby forming the upper hemisphere. Then on the lower hemisphere there is a similar hexagon. The two hexagons share sites along the equator of the sphere. Similar boundary conditions have been considered for Ising spin systems by others, e. g. Diego et. al.,[20], and Hoelbling and Lang [21], where finite, n.n. interaction, Ising spin systems on surfaces topologically equivalent to a sphere were studied.

The Fisher zeros of this 8-site cluster lie precisely on the two intersecting curves described by the above equation (3) where $J_2$ is replaced by the three-site interaction $J_3$ of the Baxter-Wu model.

A series of clusters of this form with "spherical" boundary conditions can be constructed. The upper hemisphere of the next larger cluster in the series is shown in Fig. 2b. This cluster with the spherical boundary condition is a 26-site cluster. Based on the progression from the 8-site cluster to the 26-site cluster it is easy to see how to construct a sequence of ever-larger clusters. Of course, as in any study along this line it quickly becomes extremely time consuming and memory demanding to compute the partition function. Using various symmetries and Mathematica I have calculated the exact partition function for the 8, 26, 56, and 98-site clusters. I label this sequence of clusters Series A. The complete set of Fisher zeros for this sequence of clusters are shown grouped in Fig. 3 along with the two intersecting circles given by (3). One sees graphically that all the zeros lie on the circles given. More importantly, these Fisher zeros lie precisely on the circles, as can be checked using Mathematica to whatever precision desired.

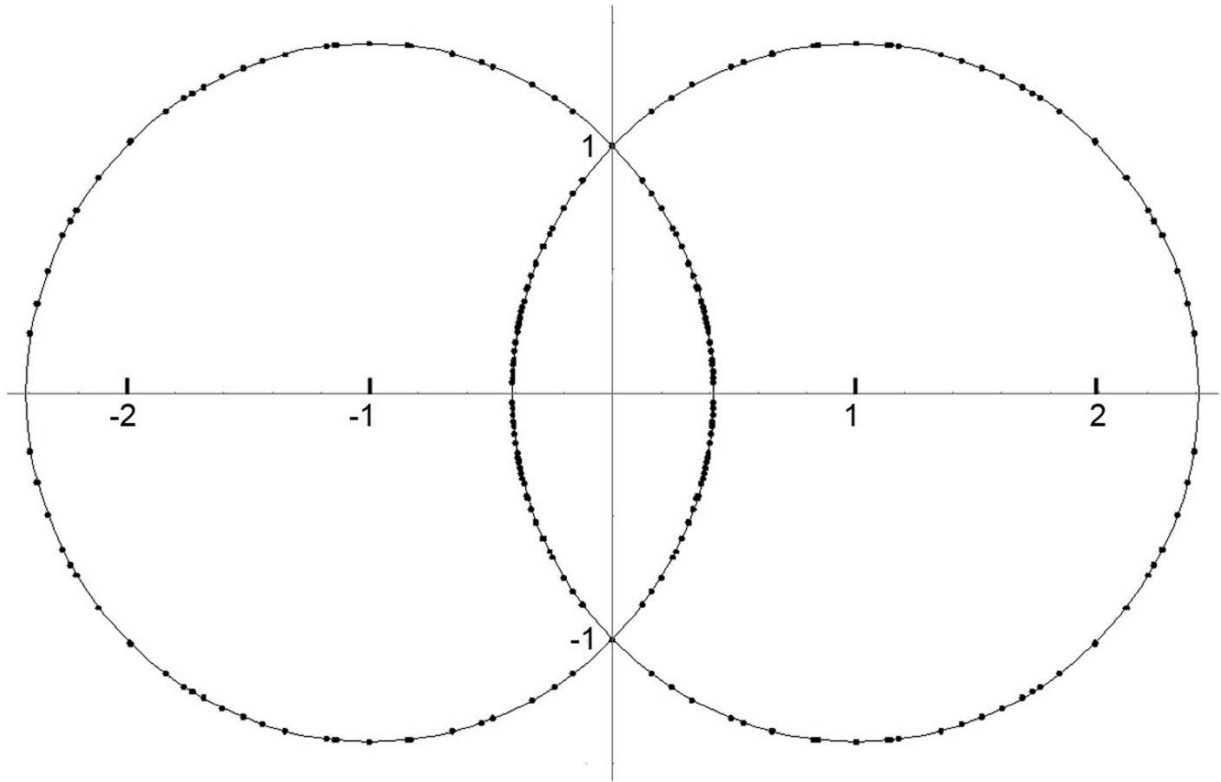

Fig. 3. The total collection of Fisher zeros from the 8, 26, 56, and 98-site clusters of Series A in the complex $e^{-2\beta J}$ plane with the two intersecting circles shown given by Eq. (3).

It is of interest to plot the Fisher zeros in complex planes other than that of $\exp[\pm 2\beta J_3]$. as the locations are special in many planes. In my initial calculations of the partition function using the variable b as written above the zeros lie on two Cassini curves, while in the complex $\exp[\pm 4\beta J_3]$ they lie on a cardioid. Brascamp and Kunz [19] point out that for their system the zeros lie on the unit circle in the complex $\sinh[2\beta J_2]$. This of course holds for the Fisher zeros of the four clusters mentioned above with spherical boundary conditions and the Baxter-Wu Hamiltonian. In the complex $\cosh[2\beta J_3]$ plane they lie on a lemniscate.

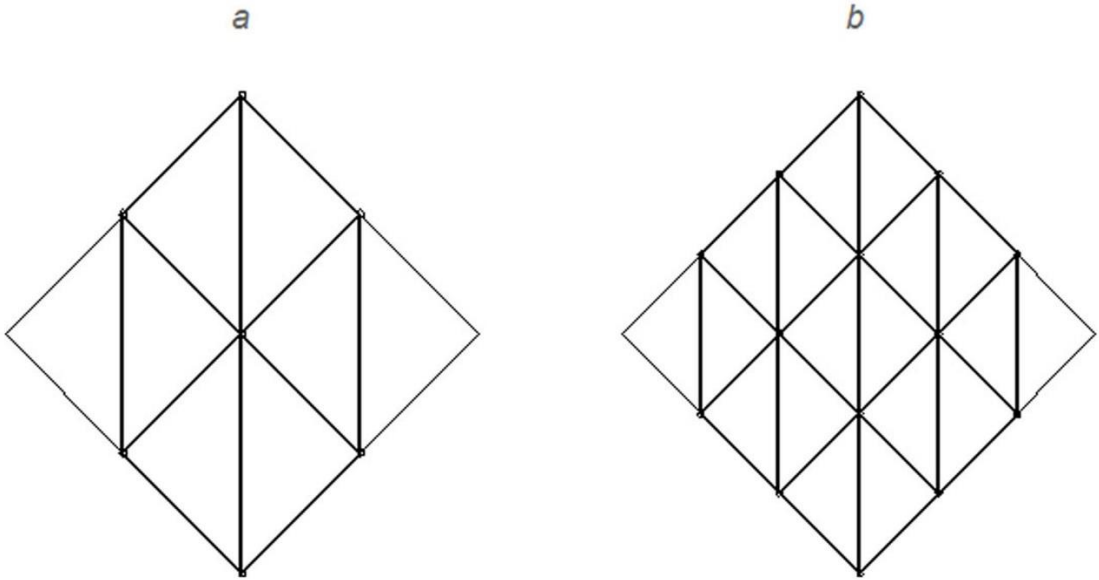

Fig. 4. The first and second clusters in the Series B. Only the bold lines are part of the clusters, not the thin lines on the left and right corners.

There is a second series of clusters, labelled Series B, where the Fisher zeros lie on the same two intersecting circles in the complex exp[±2βJ₃]. This is a series of clusters best viewed by considering the *LxL* square lattice with a diagonal along one direction forming a triangle lattice as done at the beginning of this section. The first two clusters in this sequence are illustrated in Fig. 4. Note the square clusters are clipped on two of the corners as shown. The clusters, similar to what was done in Series A, are thought of as comprising the upper half of a sphere, with an identical cluster on the lower half, and shared sites along the equator of the sphere. In this case the exact partition function of clusters of size 8, 18, 32, 50, 72, and 98-sites have been calculated. Their Fisher zeros all lie precisely on the loci as with the previous series of clusters i.e., circles, Cassini curves, intersecting circles, etc., depending on the complex plane being considered.

## IV. Critical exponents and finite size scaling.

Now with the location of the Fisher zeros for the two series of clusters with spherical boundary conditions lying on the unit circle in the complex Sinh[2βJ₃] plane it would be only reasonable to conjecture, assuming that the critical temperature of the Baxter-Wu model was not already known, that the critical temperature should be given by Sinh[2β_cJ₃] = 1. Having the solution of Baxter and Wu this is no longer a conjecture but clearly, as far as the critical temperature is concerned, the loci of the Fisher zeros shown here reflect exactly what is expected.

I now use the leading Fisher zeros for each cluster and FSS to obtain estimates of $1/\nu$, where $\nu$ is the correlation length critical exponent. Itzykson, et. al. [22] established a FSS relation between the Fisher zeros, the correlation length critical exponent $\nu$, and the critical temperature writing

$$u_i(L) - u_c = AL^{-1/\nu} \tag{4}$$

where $u = \exp[-4\beta J]$, $u_i(L)$ is the *i*-th Fisher zero for a cluster of size $L$, $u_c$ is $u$ at its critical value and $A$ is a constant and as Itzykson, et. al. point out, a complex number in general. Several authors [23, 24, 25] used this relation, or variations of it, see the following, both to examine its ability to estimate the critical temperature and to estimate $\nu$. Most studies in the past have dealt with one or more of the following three variations of (4).

$$Im[u_i(L)] \sim L^{-1/\nu}[1 + \mathcal{O}(L)] \tag{5a}$$

$$|u_i(L) - u_c| \sim L^{-1/\nu}[1 + \mathcal{O}(L)] \tag{5b}$$

$$Re[u_i(L)] - u_c \sim L^{-1/\nu}[1 + \mathcal{O}(L)] \tag{5c}$$

But as Janke and Kenna [5, 26] note, (5c) is more properly written

| | $|Re[z_i(L)] - z_c| \sim L^{-\lambda}$ | (6) |
|---|---|---|

where $1/\nu$ in (5c) is replaced with $\lambda$, a shift exponent and $z_i(L)$ is the i-th Fisher zero for some temperature variable. While Eqs. (5a), (5b), (5c), and (6) are written for general i it is generally the case, and will be the case here, that only the "leading" or "first" Fisher zeros is used. By "leading" or "first" Fisher zero is meant the zero with the minimum argument measured from the real axis.

Janke and Kenna [5,26] state the shift exponent in (6) usually equals $1/\nu$ but this is not a consequence of FSS and is not always true. In fact, they illustrate this [5, 26] by showing in the case of clusters of Ising spins, with n.n. interactions on a square lattice, and Brascamp and Kunz boundary conditions, that $\lambda = 2 = 2/\nu$. The later equality is because for their lattice spin system $\nu=1$. In the following I show that the same thing occurs with the Baxter-Wu model and Series A or B clusters, that is $\lambda = 2/\nu$. With the Baxter-Wu model $\nu = 2/3$ and thus $\lambda = 3$. While in [26] their results are analytical results, covering all sized systems, the results here are only specifically shown for the finite size clusters for which I have been able to explicitly compute the exact partition function.

Using (5a) one can obtain an estimated value for $1/\nu$ given the leading zeros from two cluster sizes using

$$\frac{1}{\nu(L,L')} = \frac{Ln\left[Imz_1(L')/Imz_1(L)\right]}{Ln\left[L'/L\right]} \quad (7)$$

Two issues must be addressed when using (7). First, unlike many previous analyses, mostly of the square lattice n.n., Ising model, where the values for *L* and *L'* are obvious, that is not the case here. I have used the square root of the number of sites in the cluster for my value of *L*. Second, while in the original derivation of the FSS scaling relation [22], u = exp[-4βJ] this is not a requirement. Janke and Kenna stated in regards to 5a, 5b, 5c, and 6 that u need only by some appropriate temperature variable. I have used for u the following, exp[-2βJ], exp[-4βJ], and Sinh[2βJ]. Previous authors have generally used exp[-2βJ], or exp[-4βJ]. There are advantages to using more than one expression for u as will be seen in the following.

TABLE I. Comparison of estimates of 1/ν for the Baxter-Wu model based on Eq. (5a).

| SERIES A | | | | SERIES B | | | |
|---|---|---|---|---|---|---|---|
| Size of systems used | Im(exp(-2βJ)) | Im(exp(-4βJ)) | Im(Sinh(2βJ)) | Size of systems used | Im(exp(-2βJ)) | Im(exp(-4βJ)) | Im(Sinh(2βJ)) |
| 8 & 26-site | 1.647503 | 1.526243 | 1.265705 | 8 & 18-site | 1.617750 | 1.466454 | 1.155446 |
| 26 & 56-sites | 1.547942 | 1.527331 | 1.466931 | 18 & 32-sites | 1.535870 | 1.491876 | 1.371021 |
| 56 & 98-sites | 1.521821 | 1.514760 | 1.493178 | 32 & 50-sites | 1.514173 | 1.494763 | 1.437516 |
| | | | | 50 & 72-sites | 1.506947 | 1.496587 | 1.465212 |
| | | | | 72 & 98-sites | 1.503884 | 1.497688 | 1.478694 |

The values of 1/ν(*L*,*L'*) obtained using (7) and the imaginary part of the leading zeros from both Series A and Series B clusters are presented in Table I. Values for the three temperature variables mentioned in the previous paragraph are given. As stated above, Baxter [8] found the value of ν to be 2/3. Hence one sees from Table I that estimates using the Fisher zeros are reasonably accurate and, except for one sequence, monotonically increase or decrease toward the exact value of Baxter. In fact, it is not unreasonable to conclude after examining the results using as the variables exp[-2βJ] and exp[-4βJ] that 1/ν must lie between 1.504 and 1. 498. In all studies the author is aware of where Fisher are used to estimate critical exponents the authors looked at results based on only one variable. The results in Table 1 show it might prove beneficial, to look at more than the results from one variable, based on the above. Based on the above, and taking the average of what appears as an upper bound and lower bound on 1/ν this approach gives 1/ν = 1.501(3). Novotny, Landau, and Swendsen [13] get $y_T$= 1.48(3) using Monte Carlo renormalization group methods and more recently Velonakis and Martinos [27] obtain ν = 0.67(4).

TABLE II. Comparison of estimates of λ for the Baxter-Wu model based on Eq. (6).

| SERIES A | | | | SERIES B | | | |
|---|---|---|---|---|---|---|---|
| Size of systems used | Re(exp(-2βJ)) | Re (exp(-4βJ)) | Re (Sinh(2βJ)) | Size of systems used | Re (exp(-2βJ)) | Re (exp(-4βJ)) | Re (Sinh(2βJ)) |
| 8 & 26-site | 3.312064 | 3.285579 | 2.833919 | 8 & 18-site | 3.256787 | 3.223863 | 2.675126 |
| 26 & 56-sites | 3.098882 | 3.094240 | 3.000481 | 18 & 32-sites | 3.078101 | 3.068250 | 2.876570 |
| 56 & 98-sites | 3.044675 | 3.043078 | 3.010272 | 32 & 50-sites | 3.031174 | 3.026798 | 2.938119 |
| | | | | 50 & 72-sites | 3.015407 | 3.013066 | 2.964877 |
| | | | | 72 & 98-sites | 3.008675 | 3.007273 | 2.978212 |

Next using (6) one can obtain estimates for λ based on the real part of the leading Fisher zeros based on two clusters. Again, as when using $Im[u_i(L)]$, I have presented the results for the cases where u = exp[-2βJ], exp[-4βJ], and sinh[2βJ]. These results can be found in Table II and as stated earlier I get accurate results similar to that found analytically, by Janke and Kenna [5,26], using Brascamp-Kunz boundary conditions. They found the shift exponent to be $\lambda = 2/\nu$. As with the results using the imaginary part of the leading zeros the series of results are monotonically decreasing or increasing toward what would appear to be a value of 3. Again, considering that one has what would seem to be upper and lower bounds one gets for the shift exponent λ = 2.99(2).

TABLE III. Comparison of estimates of 1/ν for the Baxter-Wu model based on Eq. (5b).

| SERIES A | | | | SERIES B | | | |
|---|---|---|---|---|---|---|---|
| Size of systems used | u = (exp(-2βJ)) | u = (exp(-4βJ)) | u = (Sinh(2βJ)) | Size of systems used | u = (exp(-2βJ)) | u = (exp(-4βJ)) | u = (Sinh(2βJ)) |
| 8 & 26-site | 1.451184 | 0.284902 | 0.090564 | 8 & 18-site | 1.628394 | 1.677451 | 1.337563 |
| 26 & 56-sites | 1.549441 | 1.556637 | 1.500240 | 18 & 32-sites | 1.539050 | 1.554211 | 1.438285 |
| 56 & 98-sites | 1.522337 | 1.524824 | 1.505014 | 32 & 50-sites | 1.515586 | 1.522372 | 1.469059 |
| | | | | 50 & 72-sites | 1.507703 | 1.511344 | 1.482439 |
| | | | | 72 & 98-sites | 1.504337 | 1.506520 | 1.489106 |

One final relation should be examined and that is (5b) where both the real and imaginary part of the leading zero comes into play. Once again using the leading Fisher zeros, in some appropriate variable, leads to an estimate of $1/\nu = 1.497(8)$. Results using this approach are presented in table III.

## IV. Conclusions.

After calculating the exact expression of the partition function, in terms of $\exp[\beta J_3]$ and $\exp[\beta h]$, for certain clusters of Ising spins, denoted Series A and B, governed by the Baxter-Wu Hamiltonian (1), the Fisher zeros were calculated. With the spherical boundary conditions given these clusters, the Fisher zero were shown to lie on very simple loci. Specifically, in the case of the complex $\exp[\pm 2\beta J_3]$-plane the zeros lie on two intersecting circles, just as the Fisher zeros of the n.n., square lattice, Ising model were shown to lie on two intersecting circles, when they had Brascamp-Kunz boundary conditions, and in the case of the complex $\exp[\pm 2\beta J_2]$-plane. While it seems, as with the n.n., square lattice, Ising model with Brascamp-Kunz boundary conditions, that this is true for arbitrarily large clusters, this unfortunately has not been established. In proving their result Brascamp and Kunz relied upon previous expressions [28] for the partition function and properties of their system which are currently not known for the Baxter-Wu model.

In addition to the location of the Fisher zeros, estimates for both $1/\nu$ and the shift exponent $\lambda$, defined in (6), were obtained. Two series of estimates for $1/\nu$, based on using either the imaginary part or the entire part of the leading Fisher zeros, were presented. Not surprisingly these estimates improved with an increase in the size of the clusters used, and in addition, it was seen that using variables in one complex plane gave monotonically increasing estimates while those using another variable gave monotonically decreasing estimates. Together they result in what could be conjectured to be upper and lower bounds. The estimates based on the Fisher zeros of the Series B clusters gave the more accurate results. This series using (5a) estimates resulted in an estimate of gives $1/\nu = 1.501(3)$ while using (5b) gives $1/\nu = 1.497(8)$.

In terms of the shift exponent $\lambda$, it was shown for the systems studied here, that the shift exponent is not equal to $1/\nu$, but rather $\lambda = 2/\nu$, paralleling what was found by Janke and Kenna for the n.n., square lattice, Ising model with Brascamp-Kunz boundary conditions. This should, as they state, further "resolve some hitherto puzzling features of FSS."[5]. That the Baxter-Wu model is a multi-site interaction, Ising model and not a pair interaction Ising model these results extend the class of models over which this $\lambda \neq 1/\nu$ behavior occurs.

[1] C. N. Yang and T. D. Lee, Phys Rev. **87**, 404 (1952).

[2] T. D. Lee and C. N. Yang, Rev. Rev. **87**, 410 (1952).

[3] M. E. Fischer, Boulder Lectures in Theoretical Physics, 1964 (University of Colorado Press, Boulder 1965), Vol. 7c, pg. 1.

[4] I. Bena, M. Droz, and A. Lipowski, Int. J. Mod. Phys. B **19**, 4269-4329 (2005).

[5] W. Janke and R. Kenna, Nucl. Phys. B (Proc. Suppl.) **106**, 929 (2002).

[6] R. J. Baxter, and F. Y. Wu, Phys. Rev. Letts., **31**, 1294, (1973).

[7] R. J. Baxter, and F. Y. Wu, Austr. J. Phys. **27**, 357 (1974).

[8] R. J. Baxter, Austr. J. Phys. **27**, 368 (1974).

[9] L. Onsager, Phys. Rev. **65**, 117 (1944).

[10] M. G. Watts, J. Phys. A **7**, L85, (1974).

[11] H. P. Griffiths and D. W. Wood, J. Phys. C **6**, 2533 (1973).

[12] M. P. M. den Nijs, A. M. M. Pruisken, and J. M. J. van Leeuwen, Physic **84A**, 539(1976).

[13] M. A. Novotny, D. P. Landau, and R. H. Swendsen, Phys. Rev. B **26**, 330 (1982).

[14] N. Schreiber, and J. Adler, J. Phys A: Math Gen. **38**, 7253, (2005).

[15] S. Ono, Y, Karaki, M. Suzuki and C. Kawabata, Phys. Letters **24A**, 703 (1967).

[16] S. Katsura Prog. Theor. Phys. **738**, 1415 (1967).

[17] N. A. Alves, J. R. D. de Felicio, and U. H. E. Hansmann, Int. J. of Mod. Physics C **8**, 1063 (1997).

[18] S-Y Kim, C-O Hwang, J. M. Kim, Nuclear Physics B **805**, 441 (2008).

[19] H. J. Brascamp and H. Kunz, J. Math. Phys. **15**, 65 (1974).

[20] O. Diego, J. Gonzales, and J. Salas, J. Phys. A **27**, 2965 (1994).

[21] Ch. Hoelbling, and C. B. Lang. Phys. Rev. B **54**, 3434 (1996).

[22] C. Itzykson, R. B. Pearson, & J. B. Zuber, Nuclear Physics B **220**, 415 (1983).

[23] G. Bhanot, Num. Math. **60**, 55 (1990).

[24] R. J. Creswick, Phys. Rev. E, **52**, R5735, (1995).

[25] S-Y Kim, J. Korean Phys. Soc. **62**, 214 (2013).

[26] W. Janke and R. Kenna, Phys. Rev. B **65**, 064110 (2002).

[27] I. N. Velonakis, and S. S. Martinos, Physica A **392** 2016 (2013).

[28] B. M, McCoy, and T. T. Wu, Phys. Rev. **162** 436 (1967).